\documentstyle[preprint,aps]{revtex} 
\begin{document} 
\draft

\title{ Dirac-K\" ahler approach connected to 
quantum mechanics in Grassmann space} 
\author{ N. Manko\v c Bor\v stnik} 
\address{ Department of Physics, University of 
Ljubljana, Jadranska 19,\\ 
 and  J. Stefan Institute, Jamova 39,\\ 
Ljubljana, 1111, Slovenia } 
\author{ H. B. Nielsen} 
\address{Department of Physics,  Niels Bohr Institute, 
Blegdamsvej 17,\\  
Copenhagen, DK-2100,\\ 
 and TH Division, CERN, 
CH-1211 \\ 
Geneva 23, Switzerland } 
\date{\today}

\maketitle 
 
\begin{abstract} 
We compare the way one of us got spinors out of 
fields, which are a priori antisymmetric tensor fields, to the 
Dirac-K\" ahler rewriting. Since using our Grassmann 
formulation is simple it may be useful in describing the 
Dirac-K\" ahler formulation of spinors  and in generalizing it to vector 
internal degrees of freedom and to charges. The ''cheat''  concerning 
the Lorentz transformations for spinors is the same in both 
cases and is put  
clearly forward in the Grassmann formulation. Also the 
generalizations are clearly pointed out. The discrete symmetries 
are discussed, in particular the appearance of two kinds of 
the time-reversal operators as well as the unavoidability of 
four families.  
\end{abstract} 

\section{  Introduction} 
K\" ahler\cite{kah} has shown how to pack the Dirac wave function into  
the language of differential forms in the sense that the Dirac 
equation is an equation in which  
a linear operator acts on a linear combination $u$ of $p$-forms ( p=0, 1, 
...,d; here d=dimension =4). This is the Dirac-K\" ahler  
formalism.  
 
One of us\cite{norma} has long developed an a priori rather 
different formalism  
in an attempt to unify spin and charges. In this approach the 
spin degrees of freedom come out of canonically quantizing 
certain Grassmannian odd ( position analogue in the sense of 
being on an analogue footing with $x^a$) variables 
$\theta^a$. These variables are denoted by a  
vector index $a$, and there are at first to see no spinors at 
all!    
 
One of the main purposes of the present article is to point 
out the analogy and nice relations between the two different 
ways of achieving the - almost miraculous - appearance of  
spin one half degrees of freedom in spite of starting from  
pure vectors and tensors. 
 
Of course it is a priori impossible that vectorial and tensorial 
fields ( or degrees of freedom) can be converted into spinorial 
ones without some ''cheat''.  The ''cheat'' consists really in 
exchanging one set of Lorentz transformation generators by 
another set ( which indeed means putting strongly to zero one 
type of Grassmann odd operators  fulfilling the Clifford 
algebra and  anticommuting with another type of Grassmann odd 
operators, which  also fulfill the Clifford algebra\cite{norma} ).  
 
In  fact one finds on page 512 in the K\" ahler's\cite{kah} 
article  
that there are two sets of rotation generators; one set for 
which the $u$  
field (in the K\" ahler's notation) transforms as a spinor field and 
another one for which it  
transforms as superpositions of vector and (antisymmetric ) 
tensor fields. Analogously in the approach of one of us 
the apriori Lorentz transformation generators ${\cal }S^{ab}= 
\tilde{S}^{ab} + \tilde{\tilde{S}}{ }^{ab}$ have the wave function 
transform as vectors and antisymmetric tensors, while 
$\tilde{S}^{ab}$ ( $\; = -i \frac{1}{4} [\tilde{a}^a, \tilde{a}^b]$ ) 
or $ \tilde{\tilde{S}}{ }^{ab}$ ( $\; = -i \frac{1}{4} 
[\tilde{\tilde{a}}{ }^a, \tilde{\tilde{a}}{ }^b]$ and $[,]$ means 
the commutator ) used alone are  
also possible Lorentz generators for which now the wave function 
transforms as a spinor wave function. By putting 
$\tilde{\tilde{a}}{ }^a$ ( which has the property that $ 
[\tilde{S}^{ab}, \tilde{\tilde{a}}{ }^c] = 0$  ) equal strongly 
to zero is the same as replacing ${\cal S}^{ab}$ by $ \tilde{S}^{ab}$.

In both approaches to get spinors out of vectors and 
antisymmetric tensors, as start, you get not only one but several 
copies, families, of Dirac fields. This is a fundamental feature 
in as far as these different families are connected by the 
generator parts not used: If one for instance uses 
$\tilde{S}^{ab}$ as the Lorentz generator to get spinors, then 
the not used part $\tilde{\tilde{S}}{ }^{ab}$ transforms the 
families ( of the same Grassmann character ) into each other. 
 
It will be a major content of the present article to  
bring about a dictionary relating the two formalisms  
so that one can enjoy the simplicity of one also working on the  
other one. We also shall generalize the K\" ahler operators for 
$d=4$, comment on the discrete symmetries, which in the one of 
us approach show up clearly and use the $d-4$ dimensions to 
describe spins and charges\cite{norma}.

In the following section we shall put forward the little part of 
the formalism of the work of one of us needed for the comparison 
with the Dirac-K\" ahler formalism.  
 
In the next section again   
- section 3 - we shall then  tell the (usual) Dirac-K\" ahler  
formalism as far as relevant.  
 
The comparison which should now be rather obvious is performed  
in section 4.  
 
In section 5 we shall  analyse in the two approaches  
in parallel how the remarkable finding of the Dirac equation 
inside a purely tensorial-vectorial system of fields occurs. 
 
In section 6 we shall comment on the evenness of the 
$\gamma^a$ matrices, which have to transform Grassmann odd wave 
functions into Grassmann odd wave functions. 
 
In section 7 we shall comment on discrete symmetries for either 
K\" ahler or the one of us approach,  also discussing the 
realization of the discrete symmetries pointed up clearly by 
Weinberg in his book\cite{wein} on pages 100-105. 
 
In section 8 we want to investigate how unavoidable the 
appearance of families is to this type of approaches. 
 
In section 9 we shall look at how the ideas of one of us  
of extra dimensions  generalizes the K\" ahler approach.        
 
In section 10 we discuss the Nielsen and Ninomija~\cite{nn} no 
go theorem for spinors on a lattice and a possible way out. 
 
In section 11 we shall resume and deliver concluding remarks.

\section{Dirac equations in Grassmann space}

What we can call the Manko\v c approach\cite{norma}, and which is 
the work of  
one of us, is a rather ambitious model for going beyond the  
Standard Model with say 10 ( or more) extra dimensions, but what 
we need  
for the present connection with the Dirac-K\" ahler\cite{kah} 
formalism  
is only the way in which the spin part of the Dirac particle 
fields comes about. The total number of dimensions in the  
model is ( most hopefully ) 13 +1 bosonic degrees of freedom, 
i.e. normal  
dimensions, and the same number of fermionic ones. 
 
Let us call the dimension of space-time $d$ and then the  
Dirac spinor degrees of freedom shall come from the odd 
Grassmannian variables $\theta^a, \quad a \in \{ 0,1,2,3,5, 
\cdot ,d \}$.   
 
In wanting to quantize or just to make Poisson brackets 
out of the $d$ $\theta^a$'s we have two choices 
since we could either decide to make the different $\theta^a$'s  
their own conjugate, so that one only has $d/2$ degrees of 
freedom - this is the approach of Ravndal and DiVecchia\cite{ravn} - 
or we could decide to consider the $\theta^a$'s configuration  
space variables only. In the latter case - which is the Manko\v 
c case - we have then from the $\theta^a$'s different conjugate  
variables $p^{\theta a}$.  
 
In this latter case we are entitled to write wave functions 
of the form 
\begin{equation} 
\psi (\theta^a) = \sum_{i=0,1,..,3,5,..,d} \quad \sum_{\{ a_1< 
a_2<...<a_i\}\in \{0,1,..,3,5,..,d\} } 
\alpha_{a_1, a_2,...,a_i}\theta^{a_1}  
\theta^{a_2} \cdots \theta^{a_i}. 
\label{*} 
\end{equation}  
This is the only form a function of the odd Grassmannian 
variables $\theta^a$ can take. Thus the wave function  
space here has dimension $2^d$. 
Completely analogously to usual quantum mechanics we have the 
operator for the conjugate variable $\theta^a$ to be 
\begin{equation} 
p^{\theta}_a := -i \overrightarrow{\frac{\partial}{\partial 
\theta^a }} := -i\overrightarrow{\partial}_a. 
\label{pt} 
\end{equation}  
The right arrow here just tells, that  
the derivation has to be performed from the 
left hand side. These operators 
 then obey the odd Heisenberg algebra, 
which written by means of the generalized commutators  
\begin{equation} 
\{ A, B \}: = AB - ( -1)^{n_{AB}} BA 
\label{gc} 
\end{equation} 
where  
\begin{equation} 
n_{AB}=\left\{ \begin{array}{rl} +1, \quad \hbox{if A and B have 
Grassmann odd character}\\ 0, \quad \hbox{otherwise}\end{array} \right. 
\label{gen} 
\end{equation}    
takes the form 
\begin{equation} 
\{p^{\theta a},p^{\theta b}\} = 0 = \{\theta^a, \theta^b \} , \quad 
\{ p^{\theta a}, \theta^b \} = -i \eta^{ab}. 
\label{pptt} 
\end{equation} 
Here $\eta^{ab}$ is the flat metric $\eta = diag\{1,-1,-1,...\}$.

For later use we shall define the operators  
\begin{equation} 
\tilde{a}^a := i(p^{\theta a}-i\theta^a), \quad\tilde{ 
\tilde{a}}{ }^a := 
-(p^{\theta a}+i\theta^a), 
\label{eq6} 
\end{equation} 
for which we can show that the $\tilde{a}^a$'s among themselves  
fulfill the Clifford  
algebra as do also the $\tilde{\tilde{a}}{ }^a$'s, while they 
mutually anticommute: 
\begin{equation} 
\{ \tilde{a}^a, \tilde{a}^b\} = 2\eta^{ab} = \{\tilde{\tilde{a}}{ 
}^a,  
\tilde{\tilde{a}}{ }^b\}, \quad \{\tilde 
{a}^a, \tilde{\tilde{a}}{ }^b\} = 0. 
\label{eq7} 
\end{equation} 
Note that the linear combinations (\ref{eq6}) presuppose a 
metric tensor, since otherwise only $\theta^a$ and 
$p^{\theta}{ }_a$ but not $\theta_a$ and $p^{\theta \; a}$ 
are defined.   
 
We could recognize formally 
\begin{equation} 
{\rm either} \quad \tilde{a}^a p_a|\psi> = 0, \qquad {\rm or} \quad  
\tilde{\tilde{a}}{ }^a p_a|\psi> = 0  
\label{d} 
\end{equation} 
as the Dirac-like equation, because of the  
above generalized 
commutation relations. 
Applying either the operator $\tilde{a}^a p_a$ or $\tilde{\tilde{a}}{ 
}^a p_a$ on the two equations (Eqs.(\ref{d})) we get the  
Klein-Gordon equation $p^ap_a|\psi> = 0$. 
Here of course we defined  
\begin{equation} 
p_a = i\frac{\partial}{\partial x^a}. 
\end{equation} 
However, it is rather obvious that these equations (\ref{d}) are 
not Dirac equations in the sense of the wave function 
transforming as a spinor, w.r.t. to  
the generators for the Lorentz transformations, if taken as usual 
\begin{equation} 
{\cal S}^{ab}:= \theta^a p^{\theta b} - \theta^b p^{\theta a}. 
\label{vecs} 
\end{equation} 
However it is easily seen that we can write these generators as the  
sum 
\begin{equation} 
{\cal S}^{ab} = \tilde{S}^{ab} + \tilde{\tilde{S}}{ }^{ab}, 
\label{vecsp} 
\end{equation} 
where we have defined 
\begin{equation} 
\tilde{S}^{ab} := -\frac{i}{4} [\tilde{a}^a, \tilde{a}^b], \quad 
\tilde{\tilde{S}}{ }^{ab} := -\frac{i}{4}[\tilde{\tilde{a}}{ 
}^a,\tilde{\tilde{a}}{ }^b],   
\label{S} 
\end{equation} 
with $[A,B]:=AB-BA$. One can now easily see that the solutions 
of the two equations (\ref{d}) now transform as spinors with 
respect to either $ \; \tilde{S}^{ab} \; $ or $ 
\tilde{\tilde{S}}{ }^{ab}.$  
 
It is of great importance for the ''trick'' of manipulating what we 
shall consider to be the Lorentz transformations and thus to be 
able to make the ''miraculous '' shifts of Lorentz 
representation that is the somewhat remarkable characteristic of 
the K\" ahler type of shift in formulation interpretation, 
that both - untilded, the single tilded and the double tilded -  
${\cal S}^{ab}$ obey the $d$-dimensional Lorentz generator algebra 
$\{ M^{ab}, M^{cd} \} = -i(M^{ad} \eta^{bc} + M^{bc} \eta^{ad}  
- M^{ac} \eta^{bd} - M^{bd} \eta^{ac})$, when inserted for 
$M^{ab}$.  
 
Really the ''cheat'' consist -as we shall return to - in 
replacing the Lorentz generators by the $\tilde{S}^{ab}$, say. 
This ''cheat'' means indeed that for this choice the operators  
$ \tilde{\tilde{a}}{ }^a $ have to be put strongly to zero in the 
generators of the Lorentz transformations 
(Eq.(\ref{vecs}, \ref{vecsp}, \ref{S})) 
as well as in all the other operators, representing the physical 
quantities. 
 
We shall present the one of us approach in further details in 
section 4 pointing out the similarities between this approach 
and the K\" ahler approach and generalizing the K\" ahler 
approach.  
 
\section{K\" ahler formulation of spinors }

The K\" ahler formulation\cite{kah} takes its starting point by 
considering p-forms in the $d$-dimensional space, $d = 4.$ 
Elegantly,   
the 1-forms say are defined as dual vectors to the (local) 
tangent spaces, and the higher p-forms can then be defined as  
antisymmetrized cartesian (exterior) products of the one-form 
spaces, and  
the 0-forms are the scalars; 
but we can perhaps more concretely think about the p-forms 
as formal linear combinations of the differentials of the  
coordinates $dx^a$ : 
A general linear combination of forms is then written 
\begin{equation} 
u=u_0 + u_1 +...+u_d 
\label{u} 
\end{equation} 
where the p-form is of the form 
\begin{eqnarray} 
u_p = \frac{1}{p!} \sum_{i_1,i_2,...,i_p} a_{i_1i_2...i_p}\cdot 
dx^{i_1}\wedge dx^{i_2}\wedge dx^{i_3} \wedge \cdots \wedge 
dx^{i_p} =  
\nonumber 
\end{eqnarray} 
\begin{eqnarray} 
\sum_{ i_1<i_2...<i_p}  a_{i_1i_2...i_p}\cdot 
dx^{i_1}\wedge dx^{i_2}\wedge dx^{i_3} \wedge \cdots \wedge 
dx^{i_p}.  
\label{up} 
\end{eqnarray} 
 
Then one can define both the presumably most well known exterior 
algebra denoted by the exterior product $\wedge$ and the 
Clifford product $\vee$ among the forms. The wedge product 
$\wedge$ has the property of making the product of a p-form and 
a q-form be a (p+q)-form, if a p-form and a q-form have no 
common differentials. The Clifford product $dx^a \vee$ on a 
p-form is either a $p + 1$ form, if a p-form does not include a 
one form $dx^a$, or a $p-1$ form, if a one form $dx^a$ is  
included in a p-form.

Actually  K\" ahler found how the Dirac equation  
could be written as an equation\cite{kah} ( Eq. (26.6) in the 
K\" ahler's paper) 
\begin{equation} 
-i \delta u = ( m + e\cdot \omega ) \vee u 
\label{dk1} 
\end{equation} 
where the symbol $u$ stands for a linear combination of 
$p$-forms (Eq.(\ref{u}))  and $p\in \{0,1,..,3,5,..,d\} $  
with d being the dimension  
of space-time, namely d = 4 for the K\" ahler's case. Further in the 
notation of K\" ahler the  
symbol\footnote{The 
notation in the Becher and Joos\cite{bj}  
paper is slightly different from the K\" ahler notation. The 
Becher and Joos paper  uses $d = dx^a \frac{\partial}{\partial x^a}$, 
as K\" ahler does in his paper,  but $\delta$ of K\" ahler is in 
the Becher and Joos paper replaced by $d - \delta $, which means 
that in their paper $ \delta = - e^a \frac{\delta}{\delta 
x^a}$.} $\delta$ denotes inner differentiation, which means  
the analogue of the exterior differential $d$ but with the use of 
the Clifford product $\vee$ instead of the exterior product 
$\wedge $  
\begin{equation} 
\delta u = \sum_{i=1}^3 dx^i \vee \frac{\partial u}{\partial 
x^i} - dt \vee \frac{\partial u}{\partial t} = du + \sum_{i=1}^3 
e_i \frac{\partial u}{\partial x^i} - e_i \frac{\partial 
u}{\partial t}. 
\label{dk0} 
\end{equation} 
The symbol $e \cdot \omega$ determines the coupling of the charge with 
the electromagnetic field $\omega = A_a 
dx^a, \quad a \in \{ 0,1,2,3\}$ and $m$ means the electron mass, 
the symbol $e_i$ transforms a $p$-form into $(p-1)$-form, if the 
$p$-form includes $dx^i$, otherwise it gives zero.

For a free massless particle living in a d dimensional 
space-time - this is what interests us in this paper since the 
mass term brings no new feature in the theory -  
Eq.(\ref{dk1}) can be rewritten  
in the form

\begin{equation} 
dx^a \vee p_a \;\;u = 0, \quad a = 0,1,2,3,5,...,d 
\label{vee} 
\end{equation} 
where the symbol $u$ stands again for a linear combination of 
$p$-forms ( p = 0,1,2,3,5,...,d).

That is to say that the wave function describing the state of 
the spin one half particle is packed into the exterior algebra  
function $u$. 
 
More about the K\" ahler's approach will come in  section  
4 giving the correspondence between that and the one with the 
Grassmann $\theta^a$'s, where we shall also give some 
generalizations.

\section{ Parallelism between the two approaches} 
 
We demonstrate the parallelism between the K\" ahler\cite{kah} 
and the one of us\cite{norma}  approach in  
steps, paying attention on the Becher-Joos\cite{bj} paper as 
well. First we shall treat the spin $\frac{1}{2}$ fields only, 
as K\" ahler did. We shall use the simple and  transparent 
definition of the exterior and interior product in Grassmann 
space to generalize the K\" ahler approach to two kinds of 
$\delta$ (Eq.(\ref{dk0})) operators on the space of p-forms and 
then accordingly to three kinds of the generators of the Lorentz 
transformations, two of the spinorial and one of the vectorial 
character, the first kind transforming spinor $\frac{1}{2}$ 
fields, the second one transforming the vector fields. We 
comment on the Hodge star product for both approaches, define 
the scalar product of vectors in the vector space of either p-forms 
or of polynomials of $\theta^a$'s and comment on four 
replications of the Weyl bi-spinor. We also discuss briefly the 
vector representations in both approaches.  
 
\subsection{  Dirac-K\" ahler equation and  Dirac equation in 
Grassmann  space for massless particles} 
We present here, side by side, the operators in the space of 
differential forms and in the space of polynomials of $\theta^a 
$'s. We present the exterior product 
\begin{equation} 
dx^a\; \wedge dx^b\; \wedge \cdots, \qquad \theta^a\; \theta^b 
\cdots,  
\label{ext} 
\end{equation} 
the operator of ''differentiation'' 
\begin{equation} 
-i e^a, \qquad p^{\theta a} = - i \overrightarrow{\partial^a} = 
-i \overrightarrow{\frac{\partial}{\partial \theta_a}} 
\label{dif} 
\end{equation} 
and the two superpositions of the above operators 
\begin{eqnarray} 
dx^a \;\tilde{\vee}: = dx^a \wedge + \; e^a, \quad \quad 
\tilde{a}^a:= i(p^{\theta a} - i \theta^a), 
\nonumber 
\end{eqnarray} 
\begin{eqnarray} 
dx^a \;\tilde{\tilde{\vee}}: = i(dx^a \wedge - \; e^a), \quad \quad 
\tilde{\tilde{a}}{ }^a:= -(p^{\theta a} + i \theta^a). 
\label{att} 
\end{eqnarray} 
Here $\tilde{\vee}$  stays instead of $\vee$ of 
Eq.(\ref{dk1}),  
used by K\" ahler. Introducing the notation with $\tilde{ }$ and 
$\tilde{\tilde{ }}$ we not only point out  the similarities between 
the two approaches but also the two possibilities for the 
Clifford product - only one of them used by K\" ahler. Both 
$\tilde{\vee} $ and $\tilde{\tilde{\vee}} $  
 are Clifford products on p-forms, while 
$\tilde{a}^a$ $\tilde{\tilde{a}}{ }^a$ are the corresponding 
linear operators operating on the space of polynomials of 
$\theta^a$'s.  
One easily finds the commutation relations, if  for 
both approaches the generalized form of commutators, presented 
in Eq.(\ref{gen}), are understood  
\begin{eqnarray} 
\{ dx^a \;\tilde{\vee},\; dx^b \; \tilde{\vee} \} = 2 
\eta^{ab}, \qquad  
\{ \tilde{a}^a, \tilde{a}^b \}  = 2 \eta^{ab} 
\nonumber 
\end{eqnarray} 
\begin{eqnarray} 
\{ dx^a \;\tilde{\tilde{\vee}},\; dx^b \; \tilde{\tilde{\vee}} 
\} = 2 \eta^{ab}, \qquad  
\{ \tilde{\tilde{a}}{ }^a, \tilde{\tilde{a}}{ }^b \} = 2 
\eta^{ab}.  
\label{cat} 
\end{eqnarray} 
Here $\eta^{ab}$ is the metric of  space-time. 
 
The vacuum state $ |\; > $ is defined as 
\begin{eqnarray} 
dx^a\;\tilde{\vee} \;|\; > = dx^a \wedge, \qquad \tilde{a}^a \;|\; > = 
\theta^a,  
\nonumber 
\end{eqnarray} 
\begin{eqnarray} 
dx^a\;\tilde{\tilde{\vee}} \;|\; > = dx^a \wedge, \qquad 
\tilde{\tilde{a}}{ }^a\; |\; > = \theta^a. 
\label{vac} 
\end{eqnarray} 
Now we can define the Dirac-like equations for both approaches: 
\begin{eqnarray} 
dx^a\; \tilde{\vee}\; p_a\; u = 0, \quad \quad 
\tilde{a}^a\; p_a\; \psi(\theta^a) = 0, 
\nonumber 
\end{eqnarray} 
\begin{eqnarray} 
dx^a\; \tilde{\tilde{\vee}}\; p_a\; u = 0, \quad 
\quad\tilde{\tilde{a}}{ }^a\; p_a\; \psi(\theta^a) = 0. 
\label{dkt} 
\end{eqnarray} 

Since $ \{ e^a , dx^b \; \wedge \} = \eta^{ab}$ and $ \{e^a , e^b 
 \} = 0 = \{dx^a \; \wedge, dx^b \; \wedge 
 \} = 0$, while $\{ -i p^{\theta a},  
\theta^b \} = \eta ^{ab}$ and $\{i p^{\theta a}, i p^{\theta b} \} 
= 0 = \{ \theta^a, \theta^b \}$, it is obvious that $ e^a  
 $ plays in the p-form formalism the role of the 
derivative with respect to a differential $1-$form, similarly as 
$ip^{\theta a}$ does with respect to a Grassmann coordinate.

Taking into account the above definitions, one easily finds that 
\begin{eqnarray} 
dx^a \; \tilde{ \vee}\; p_a\;\; dx^b\; \tilde{\vee}\; 
p_b \;u = p^a\; p_a \; u = 0,  
\qquad \tilde{a}^a\; p_a \;\; 
\tilde{a}^b\; p_b\; \psi(\theta^b) = p^a \; p_a\; \psi(\theta^b) 
= 0. 
\nonumber 
\end{eqnarray} 
\begin{eqnarray} 
dx^a \; \tilde{\tilde{\vee}}\; p_a\;\; dx^b\; \tilde{\tilde{\vee}}\; 
p_b \;u = p^a\; p_a \; u = 0,  
\qquad \tilde{\tilde{a}}{ }^a\; p_a \;\; 
\tilde{\tilde{a}}^b\; p_b\; \psi(\theta^b) = p^a \; p_a\; 
\psi(\theta^b) = 0.  
\label{kg} 
\end{eqnarray} 
 
Both vectors, the $u$, which are the superpositions of 
differential p-forms and the $\psi(\theta^a)$, which are polynomials in 
$\; \theta^a$'s are defined in a similar way (Eqs.(\ref{*} 
,\ref{up})), as 
we shall point out in the following subsection.

We see that either $dx^a \; \tilde{ \vee}\; p_a = 0\; $ or 
$\; dx^a\; \tilde{\tilde{ \vee}} \; p_a = 0 ,$ similarly as either 
$\tilde{a}^a \;p_a = 0\; 
$ or $\; \tilde{\tilde{a}}{ }^a \; p_a = 0 $ can represent the 
Dirac-like equation.

\subsection{Vector space of  two approaches} 
The superpositions of p-forms  
on which the Dirac-K\" ahler equation is defined  
and the superpositions of polynomials in 
Grassmann space, on which the Dirac-like equations are defined, are  
\begin{eqnarray} 
&u& = \sum_{i=0,1,2,3,5,..,d} \quad \sum_{a_1<a_2<...<a_i\in\{ 
0,1,2,3,5,..,d\}}  
a_{a_1a_2...a_p}\cdot  
dx^{a_1}\wedge dx^{a_2}\wedge dx^{a_3} \wedge \cdots \wedge 
dx^{a_i}, 
\nonumber 
\end{eqnarray} 
\begin{eqnarray} 
&\psi (\theta^a)& = \sum_{i=0,d} \quad \sum_{ a_1< 
a_2<...<a_i \in \{0,1,2,3,5,...,d\} } \alpha_{a_1 a_2...a_i} \cdot 
\theta^{a_1} \theta^{a_2} \cdots \theta^{a_i}. 
\label{funv} 
\end{eqnarray} 
The coefficients $\alpha_{a_1,a_2,..,a_i}$ depend on coordinates 
$x^a$ in both cases and are antisymmetric tensors of the rank 
$i$ with respect to indices $a_k \in \{a_1,..,a_i\}$. The 
vector space is in both cases $16$ dimensional.

\subsection{ Dirac $ \gamma^a$ -like operators} 
Both, $dx^a \; \tilde{\vee}$ and $dx^a \; \tilde{\tilde{\vee}} $ 
define the algebra of the $\gamma^a$ matrices  
and so they do both $\tilde{a}^a $ and $ \tilde{\tilde{a}}{ 
}^a$. One would thus be tempted to identify 
\begin{equation} 
\gamma_{ \hbox{naive} }^a := dx^a \; \tilde{\vee},\qquad \hbox{ or }  
\qquad \gamma_{ \hbox{naive} }^a :=  \tilde{a}^a. 
\label{naive} 
\end{equation} 
 
But there is a large freedom in defining what to identify with 
the gamma-matrices, because except when using $\gamma^0$ as 
a parity operation you have an even number of gamma matrices  
occurring in the physical applications. Then you may multiply  
all the gamma matrices by some factor provided it does not 
disturb their algebra nor their even products. 
We shall comment this point in section 6. 
 
\subsubsection{Problem of  statistics of  gamma-matrices} 
This freedom might be used to solve, what seems a problem: 

Having an odd Grassmann character, 
neither $\tilde{a}^a$ nor 
$\tilde{\tilde{a}}{ }^a$ should be recognized as the Dirac 
$\gamma^a$ operators, since they would change, when operating on 
polynomials of $\theta^a$, polynomials of an odd Grassmann 
character to polynomials of an even Grassmann character. One 
would, however, expect - since  Grassmann odd fields  
second quantize to fermions, while Grassmann even fields  
second quantize to bosons -  that the $\gamma^a$ operators do not 
change the Grassmann character of  wave functions. One can 
notice, that similarly to the Grassmann case, also the two 
types of the Clifford products defined on p-forms, change the 
oddness or the evenness of the p-forms: an even p-form, $p=2n$, is 
changed by either $dx^a \; \tilde{\vee}$ or $dx^a \;  
\tilde{\tilde{\vee}} $ to an odd $q$-form, with either $q = p+1$, if 
$dx^a $ is not included in a p-form, or  $q = p - 1$, if $dx^a$ is 
included in a $p$-form, while  an odd p-form, $ p = 2n + 1 
$, is changed to an  even $p + 1$-form or $p - 1$-form.  
 
\subsubsection{ First solution to gamma-matrix statistics 
problem}  
We 
shall later therefore propose that accordingly  
\begin{equation} 
{\rm either} \quad \tilde{\gamma}^a: = i \; dx^0 \; 
\tilde{\tilde{\vee}} \; dx^a \; \tilde{\vee}, \quad {\rm or} \qquad 
\tilde{\gamma}^a = i \; \tilde{\tilde{a}}{ }^0 \; \tilde{a}^a  
\label{eq25} 
\end{equation} 
are recognized as the Dirac $\gamma^a$ operators operating on 
the space of $p$-forms or polynomials of $\theta^a$'s, 
respectively, since they both  
have an even Grassmann character and they both fulfill the 
Clifford algebra 
\begin{equation} 
\{ \tilde{\gamma}^a, \tilde{\gamma}^b \} = 2 \eta^{ab}. 
\label{gamat} 
\end{equation} 
Of course, the role of $(\;\tilde{ }\;)$ and $(\;\tilde{\tilde{ 
}}\;) $ can in either 
the K\" ahler case or the case of polynomials in Grassmann space, be 
exchanged. 
 
Whether we define the gamma-matrices by (\ref{eq25}) or 
(\ref{naive}) makes only a difference for an odd products of 
gamma-matrices, but for applications such as construction of  
currents $\bar{\psi}\gamma^a\psi$ or for the Lorentz generators  
on the spinors $-i \frac{1}{4}\; [\gamma^a,\gamma^b]$ only products of even 
numbers of gamma-matrices occur, except for the 
parity representation on the Dirac fields, where the 
$\gamma^0$-matrix is used alone. This $\gamma^0$-matrix has to 
simulate the parity reflection which is either  
\begin{equation} 
\vec{dx}\rightarrow -\vec{dx}, \quad {\rm or} \quad 
\vec{\theta}\rightarrow - \vec{\theta}. 
\label{parity} 
\end{equation} 
The ''ugly'' gamma-matrix identifications (\ref{eq25}) indeed 
perform this operation. And as long as the physical applications 
are the ones just mentioned - and that should be sufficient - 
the choice (\ref{eq25}) is satisfactory: Living in the Grassmann 
odd part of the Hilbert space, we don't move into the Grassmann 
even part of it.  
 
The canonical quantization of Grassmann odd fields, that is the 
procedure with the Hamiltonian and the Poisson brackets, then 
automatically assures the anticommuting relations between the 
operators of the fermionic fields. 
 
\subsubsection{ Solution by  redefinition of oddness} 
The simplest solution to the problem with the evenness and 
oddness is  
to use the ''naive'' gamma-matrix identifications (\ref{naive}) 
and simply ignore that the even-odd-ness does not match. 
This is what K\" ahler did, we can say, in as far as he  
did not really identify the even-odd-ness of the p-forms 
with the statistics of Dirac fields. If one - along the  
lines of the Becher's and Joos's  paper (\cite{bj}) - will make a 
second quantized theory  
based on the K\" ahler trick one  does not proceed by insisting 
on 
taking p-forms to be fermionic only when p is odd.   
Becher and Joos take all the forms as fermion fields and assume 
then anticommuting relations for operators of fields.  
This simplest solution can thus be claimed to be the one  
applied by K\" ahler and used by Becher and Joos: They simply do 
not dream about  
in advance postulating that the p-forms should necessarily be 
taken to be boson or fermion fields depending on whether  
p is  even or odd. 
 
It is only when one as one of us in her model has the 
requirement of canonical quantization saying that the  
$\theta^a$'s should be Grassmann odd objects, which indeed they are, 
that the problem occurs.

\subsection{ Generators of  Lorentz transformations} 
Again, we are presenting the generators of the Lorentz 
transformations of spinors for both approaches 
\begin{equation} 
M^{ab} = L^{ab} + {\cal S}^{ab}, \qquad L^{ab} = x^a p^b - x^b p^a, 
\end{equation} 
differing among themselves in the definition of 
${\cal S}^{ab}$ only, which define  
the generators of the Lorentz transformations in the internal 
space, that is in the space of $p$-forms or polynomials of 
$\theta^a$'s, respectively. 
While K\" ahler suggested the definition  
\begin{equation} 
{\cal S}^{ab} = dx^a \wedge dx^b, \quad {\cal S}^{ab} u = \frac{1}{2} 
( (dx^a \wedge dx^b) \vee u - u \vee ( dx^a \wedge dx^b )), 
\end{equation}  
in the Grassmann case~\cite{norma} the operator ${\cal S}^{ab} $ 
is one of the  
two generators defined above ( Eq. (\ref{S})), that is  
\begin{equation} 
{\rm either} \quad {\cal S}^{ab} = \tilde{S}^{ab} = 
-\frac{i}{4} [\tilde{a}^a, \tilde{a}^b] = -\frac{i}{4} 
[\tilde{\gamma}^a, \tilde{\gamma}^b] , \quad {\rm or} \qquad  
{\cal S}^{ab} = \tilde{\tilde{S}}{ }^{ab}  = 
-\frac{i}{4} [\tilde{\tilde{a}}{ }^a, \tilde{\tilde{a}}{ }^b]. 
\label{eq29} 
\end{equation}

One further finds  
\begin{equation} 
[\tilde{S}^{ab}, \tilde{a}^c] = i(\eta^{ac} \tilde{a}^b - 
\eta^{bc} \tilde{a}^a), \qquad  
[\tilde{\tilde{S}}{ }^{ab}, \tilde{\tilde{a}}{ }^c] = 
i(\eta^{ac} \tilde{\tilde{a}}{ }^b - 
\eta^{bc} \tilde{\tilde{a}}{ }^a), \quad {\rm while} \quad  
[\tilde{S}^{ab}, \tilde{\tilde{a}}{ }^c] = 0 = 
[\tilde{\tilde{S}}{ }^{ab}, \tilde{a}^c].   
\end{equation} 
  
One can also in the K\" ahler case define two kinds of 
the Lorentz generators, which operate on the internal space of 
p-forms,  
according to two kinds of the Clifford products, presented above. 
Following the definitions in the one of us~\cite{norma} 
approach, one can  
write the ${\cal S}^{ab}$ for the K\" ahler case  
\begin{equation} 
{\rm either} \;\; \tilde{{\cal S}}^{ab} = -\frac{i}{4} [dx^a \; \wedge + 
\;e^a, \; dx^b\; \wedge + \; 
e^b ] =  -\frac{i}{4} [\tilde{\gamma}^a, \tilde{\gamma}^b], 
\quad {\rm or} \quad \tilde{\tilde{\cal S}}{ }^{ab} = 
\frac{i}{4} [dx^a\; \wedge -\; e^a  
,\; dx^b \; \wedge - \;e^b ]. 
\label{eq31} 
\end{equation} 
Not only are in this 
case the similarities  
between the two approaches  more transparent,  also the 
definition of the generators of the Lorentz transformations in 
the space of p-forms simplifies very much.

One further finds for the spinorial case 
\begin{equation} 
[M^{ab}, \tilde{\gamma}^a \;\; p_a] = 0, \quad {\rm for } \quad M^{ab} 
= L^{ab} + \tilde{S}^{ab}, 
\end{equation} 
which demonstrates that the total angular momentum for a free 
massless particle  is conserved. 
The above equation is true for both approaches and  
 the generators of the Lorentz transformations $M^{ab}$  
fulfill the Lorentz algebra in both cases. 
 
In addition, the operators of the Lorentz transformations with the 
vectorial character can also be defined for both approaches in 
an equivalent way, that is as a sum  of the two operators of 
the spinorial character 
\begin{equation} 
{\cal S}^{ab} = \tilde{S}^{ab} + \tilde{\tilde{S}}{ }^{ab} = -i 
( dx^a \wedge e^b  - dx^b \wedge e^a  ), \qquad {\cal 
S}^{ab} = \tilde{S}^{ab} + \tilde{\tilde{S}}{ }^{ab} = \theta^a  
p^{\theta b} - \theta^b p^{\theta a}, 
\label{vecsk} 
\end{equation} 
which again fulfill the Lorentz algebra. The operator 
$ {\cal S} ^{ab} = -i (dx^a \wedge e^b  - dx^b \wedge e^a),  $ 
if being applied on differential p-forms, transforms vectors 
into vectors, correspondingly $ {\cal S}^{ab} = \theta^a  
p^{\theta b} - \theta^b p^{\theta a} $, if being applied to 
polynomial of $\theta^a$'s transforms vectors into 
vectors~\cite{norma}.  
 
Elements of the Lorentz group can be written for both 
approaches, for either spinorial or vectorial kind of the 
generators as 
\begin{equation} 
U = e^{- \omega_{ab} M^{ab}}, 
\end{equation} 
where $\omega_{ab} $ are parameters of the group. If $M^{ab}$ 
are equal to either $L^{ab} + \tilde{S}^{ab}$ or $L^{ab} + 
\tilde{\tilde{S}}{ }^{ab}$, the period of transformations is $4 
\pi $ either in the  
space of differential forms or in the Grassmann space, 
demonstrating the spinorial character of the operator. If 
$M^{ab}$ is the sum of $ L^{ab }$ and $\tilde{S}^{ab} + 
\tilde{\tilde{S}}{ }^{ab}$, the period of transformation is $2  
\pi$, manifesting  
the vectorial character of the operator.

\subsection{  Hodge star product} 
In the way how we have defined the operators in the space of 
$p$-form, the definition of the "Hodge star" operator, defined by 
K\" ahler  
working in the space of p-forms and the space of 
$\theta^a$ polynomials, will be  respectively 
\begin{equation} 
\tilde{\Gamma} = \quad i \prod_{a=0,1,2,3,5,..,d} \tilde{\gamma}^a,  
\end{equation} 
with $\tilde{\gamma}^a $ equal to either $ dx^0 \;\; \tilde{\tilde 
{\vee}} \;\; dx^a \;\; \tilde{\vee}$ in  
the K\" ahler case, or to $ i\;\;\tilde{\tilde{a}}{ }^0 \;\; 
\tilde{a}^a $ in  
the one of us~\cite{norma} approach.  For an even d the factor 
with double tilde ( $\; \tilde{\tilde{ }}\;$ ) can be in both cases 
omitted ( $\tilde{\Gamma} = {\rm either} \;\;i\; \prod_a \; 
\tilde{a}^a \;\; {\rm or } \;\; i\; \prod_a \; dx^a \; 
\tilde{\vee} \;\;  $,  
 Again we could distinguish the operators $\tilde{\Gamma}$ and 
$\tilde{\tilde{\Gamma}}$ in both cases, according to  
 the elements, which define the Casimir ). 
It follows that 
\begin{equation} 
\frac{1}{2} (1 \pm \tilde{\Gamma}) 
\end{equation} 
are the two operators, which when being applied on wave 
functions defined either on p-forms or on polynomials in  
Grassmann space, project out the left or right handed component, 
respectively.   
 
One easily recognizes that when being applied on a 
vacuum state $| >$, the operator $\tilde{\Gamma}$ behaves as a 
"hodge star" product, since one finds for $d$ even 
\begin{equation} 
 -i\tilde{\Gamma} \;\; |\; > = dx^0 \wedge dx^1 
\wedge....\wedge dx^d, \qquad   
\quad -i\tilde{\Gamma} \;\; |\; > = \theta^0 \theta^1...\theta^d. 
\end{equation}

\subsection{ Scalar product} 
In the Manko\v c's approach~\cite{norma} the scalar product between 
the two functions $\psi^{(1)}(\theta^a)$ and $\psi^{(2)}(\theta^a)$ is 
defined as follows  
\begin{equation} 
<\psi^{(1)}|\psi^{(2)}> = \int d^d \theta \;\; ( \omega \;\; 
\psi^{(1)}(\theta^a))^{*} \;\; \psi^{(2)}(\theta^a).  
\label{scalpt} 
\end{equation} 
Here $\omega$ is the weight function 
\begin{equation} 
\omega = \prod_{ i = 0,1,2,3,5,..,d} \qquad (\theta^i + 
\overrightarrow{\partial^i})  
\end{equation} 
which operates on  the first function,  
 $\psi^{(1)}$, only while  
\begin{equation} 
\int d \theta^a = 0, \quad \int d \theta^a \theta^a = 1, \quad a 
= 0,1,2,3,5,..,d,  
\end{equation} 
no summation over repeated index is meant 
and 
\begin{equation} 
\int d^d \theta \;\theta^0 \theta^1 \theta^2 \theta^3 \theta^5 
... \theta^d = 1, \quad  
d^d \theta = d \theta^d ... d \theta^5 d \theta^3 d \theta^2 d 
\theta^1 d \theta^0.  
\end{equation}  
Since $\theta^{a*}  = \theta^a $, $\;{ }^* $ means the complex 
conjugation and $\;{ }^+ $ means the hermitian conjugation, then 
with respect to the  
above defined  
scalar product the operator  $\theta^{a+} = - \eta^{aa}  
\overrightarrow{\partial}^a, $  $ 
\overrightarrow{\partial}^{a+} = -\theta^{a}  \eta^{aa}, $  
while $\tilde{a}^{a+} = - \eta^{aa} \tilde{a}^a$ and  
$\tilde{\tilde{a}}{ }^{a+} = - \eta^{aa} \tilde{\tilde{a}}{ 
}^a$. Again no summation over repeated index is performed. 
Accordingly the operators of the Lorentz transformations  
of spinorial character are self-adjoint (if $a\neq 0$ and $b 
\neq 0$) or anti-self-adjoint ( if a = 0 or b = 0 ).  
 
According to Eqs.(\ref{scalpt}, \ref{funv}) the scalar product of two 
functions $ \psi^{(1)}(\theta^a) \; $ and $\;\psi^{(2)}(\theta^a) 
\;) $ can be written as follows 
\begin{equation} 
<\psi^{(1)}|\psi^{(2)}> \; = \; \sum_{0,d}\;\; 
\sum_{\alpha_1<\alpha_2<..<\alpha_i}  
\;\alpha^{(1)*}_{\alpha_1 \alpha_2..\alpha_i}\;\cdot\; 
\alpha^{(2)}_{\alpha_1 \alpha_2..\alpha_i} 
\label{grasscal} 
\end{equation} 
in complete analogy with the usual definition of scalar products 
in ordinary space.    
K\" ahler~\cite{kah} defined in Eq. (15.11) and on page 519 the 
scalar product of two 
superpositions of  
p-forms $u^{(1)}\;$ and $u^{(2)}$ as follows 
\begin{equation} 
<u^{(1)}|u^{(2)}>\; =\; \sum_{0,d}\;\; 
\sum_{\alpha_1<\alpha_2<..<\alpha_i}  
\;\alpha^{(1)}_{\alpha_1 \alpha_2..\alpha_i}\;\cdot\; 
\alpha^{(2)}_{\alpha_1 \alpha_2..\alpha_i}, 
\label{kahscal} 
\end{equation} 
which ( for real coefficients $\alpha^{(k)}_{\alpha_1 
\alpha_2..\alpha_i}\;, k =  
1,2\; $) agrees with Eq.(\ref{grasscal}).

\subsection{ Four families of solutions in  K\" ahler 
or in   approach in Grassmann space} 
We shall limit ourselves in $d=4$ and in spinorial case ( as 
indeed K\" ahler did). The representations for higher d, 
analyzed with respect to the groups $SO(1,3) \times SU(3) \times 
SU(2) \times U(1)$, and some other groups, in Grassmann space are 
( only for   
Grassmann even part of the space belonging to the groups which 
does not include $SO(1,3)$) presented in ref.\cite{norma}. 
 
 In the case of $d = 4$ 
 one may arrange the space of $2^d $ vectors into four times 
two Weyl spinors, one left ( $<\;\tilde{\Gamma}^{(4)}> = - 1$) and 
one right ( $<\; \tilde{\Gamma}^{(4)}> = 1$) handed. We are 
presenting this vectors, which are at the same time the 
eigenvectors of  $\tilde{S}^{12}$ and  $\tilde{S}^{03}$, 
as polynomials of $\theta^m$'s, $m \in (0,1,2,3)$. The two Weyl 
vectors are connected by the operation of $\tilde{\gamma}^m$ 
operators (Eq.(\ref{eq25})).

Taking into account that $\tilde{a}^a | \;> = \theta^a $, where 
$| \;>$ is the vacuum state (Eq.(\ref{vac})), we find 
 
\begin{center} 
\begin{tabular}{|r|r||c||r|r|r|} 
\hline 
a&i&$^a\psi_i(\theta)$&$\tilde{S}^{12}$&$\tilde{S}^{03}$&$ 
\tilde{\Gamma}^{(4)}$\\ 
\hline\hline 
1&1&$\frac{1}{2}(\tilde{a}^1 - i \tilde{a}^2) (\tilde{a}^0 - 
\tilde{a}^3)$& $-\frac{1}{2}$& $\frac{i}{2}$& -1\\ 
\hline  
1&2&$-\frac{1}{2}(1 +i\tilde{a}^1 \tilde{a}^2) (1-\tilde{a}^0  
\tilde{a}^3)$& $\frac{1}{2}$& $-\frac{i}{2}$& -1\\ 
\hline  
2&1&$\frac{1}{2}(\tilde{a}^1 - i \tilde{a}^2) (\tilde{a}^0 + 
\tilde{a}^3)$& $-\frac{1}{2}$& $-\frac{i}{2}$& 1\\ 
\hline  
2&2&$-\frac{1}{2}(1 +i\tilde{a}^1 \tilde{a}^2) (1+\tilde{a}^0  
\tilde{a}^3)$& $\frac{1}{2}$& $\frac{i}{2}$& 1\\ 
\hline\hline  
3&1&$\frac{1}{2}(\tilde{a}^1 - i \tilde{a}^2) (1-\tilde{a}^0  
\tilde{a}^3)$& $-\frac{1}{2}$& $-\frac{i}{2}$& 1\\ 
\hline  
3&2&$-\frac{1}{2}(1 +i\tilde{a}^1 \tilde{a}^2) (\tilde{a}^0 - 
\tilde{a}^3)$& $\frac{1}{2}$& $\frac{i}{2}$& 1\\ 
\hline  
4&1&$\frac{1}{2}(\tilde{a}^1 - i \tilde{a}^2) (1 +\tilde{a}^0  
\tilde{a}^3)$& $-\frac{1}{2}$& $\frac{i}{2}$& -1\\ 
\hline  
4&2&$-\frac{1}{2}(1 +i\tilde{a}^1 \tilde{a}^2) (\tilde{a}^0 +  
\tilde{a}^3)$& $\frac{1}{2}$& $-\frac{i}{2}$& -1\\ 
\hline\hline  
5&1&$\frac{1}{2}(1 -i\tilde{a}^1  \tilde{a}^2) (\tilde{a}^0 - 
\tilde{a}^3)$& $-\frac{1}{2}$& $\frac{i}{2}$& -1\\ 
\hline  
5&2&$-\frac{1}{2}(\tilde{a}^1 +i \tilde{a}^2) (1-\tilde{a}^0  
\tilde{a}^3)$& $\frac{1}{2}$& $-\frac{i}{2}$& -1\\ 
\hline  
6&1&$\frac{1}{2}(1 -i\tilde{a}^1  \tilde{a}^2) (\tilde{a}^0 + 
\tilde{a}^3)$& $-\frac{1}{2}$& $-\frac{i}{2}$& 1\\ 
\hline  
6&2&$-\frac{1}{2}(\tilde{a}^1 +i\tilde{a}^2) (1+\tilde{a}^0  
\tilde{a}^3)$& $\frac{1}{2}$& $\frac{i}{2}$& 1\\ 
\hline\hline  
7&1&$\frac{1}{2}(1 -i\tilde{a}^1 \tilde{a}^2) (1-\tilde{a}^0  
\tilde{a}^3)$& $-\frac{1}{2}$& $-\frac{i}{2}$& 1\\ 
\hline  
7&2&$-\frac{1}{2}(\tilde{a}^1 + i\tilde{a}^2) (\tilde{a}^0 - 
\tilde{a}^3)$& $\frac{1}{2}$& $\frac{i}{2}$& 1\\ 
\hline  
8&1&$\frac{1}{2}(1 -i\tilde{a}^1 \tilde{a}^2) (1 +\tilde{a}^0  
\tilde{a}^3)$& $-\frac{1}{2}$& $\frac{i}{2}$& -1\\ 
\hline  
8&2&$-\frac{1}{2}(\tilde{a}^1  +i\tilde{a}^2) (\tilde{a}^0 +  
\tilde{a}^3)$& $\frac{1}{2}$& $-\frac{i}{2}$& -1\\ 
\hline\hline  
\end{tabular} 
\end{center} 
Table I.-Irreducible representations of the two subgroups 
$SU(2) \times SU(2)$ of the group $SO(1,3)$ as defined by the 
generators of the spinorial character $\tilde{S}^{12}, 
\tilde{S}^{03}$ and the operator of handedness 
$\tilde{\Gamma}^{(4)} $. The four copies of the Weyl bispinors 
have either an odd or an even Grassmann character. The generators 
$\tilde{\tilde{S}}{ }^{mn}$, $m,n \in (0,1,2,3)$, transform the 
two copies  
of the same Grassmann character one into another.   
 
Similarly also the K\" ahler spinors can be arranged into four 
copies. We find them by only replacing in Table I. $\tilde{a}^a$ 
by $dx^a \tilde{\vee}$. We shall discuss this point also in the 
next section.

\subsection{Vector representations of  group $SO(1,3)$} 
Analysing the irreducible representations of the group $SO(1,3)$ 
in analogy with the spinor case but taking into account the 
generator of the Lorentz transformations of the vector type 
(Eqs.(\ref{vecs}, \ref{vecsk})) one finds~\cite{norma} for d = 4 
two scalars  
( a scalar and  
a pseudo scalar), two three vectors (in the $SU(2) \times SU(2) 
$ representation of $SO(1,3)$ usually denoted by $(1,0) $ and $(0,1)$ 
representation, respectively, with $<\Gamma^{(4)}>$ equal to $\pm 
1$ ) and two four 
vectors ( in the $SU(2) \times SU(2) 
$ representation of $SO(1,3)$ both denoted by 
$(\frac{1}{2},\frac{1}{2}) $ and differing among themselves in the 
Grassmann character ) all of which are 
eigenvectors of  
$S^{(4)2 } = \frac{1}{2} S^{ab}S_{ab}, \;\; \Gamma^{(4)} = 
i\frac{(-2i)^2}{4!} \epsilon_{abcd} {\cal S}^{ab} {\cal S}^{cd}, 
\;\;{\cal S}^{12}\;\; {\rm and} \;\; {\cal }S^{03}$. Using 
Eq.(\ref{vecsk})  
and analyzing the vector  
space of p-forms in an analogous way as the space of the 
Grassmann polynomials, one finds the same kind of 
representations also in the K\" ahler case.  
 
Both, in the spinor case and in the vector case one has $2^4$ 
dimensional vector space.

\section{ Appearance of Spinors} 
One may quite strongly wonder about how it is at all possible  
that there appear the Dirac equation - usually being an equation 
for a {\em spinor} field - out of models with only scalar, 
vector and tensor objects! Immediately one would say that it 
is of course sheer impossible to construct spinors such as Dirac 
fields out of the integer spin objects such as the differential 
one forms and their external products or of  the $\theta^a$'s 
and their products $\theta^a\theta^b\cdots \theta^c$.  
 
Let us say immediately that it also only can be done by a 
''cheat''. This ''cheat''  
really consists in {\em replacing} the Lorentz  
transformation concept ( including rotation concept)  
by exchanging the Lorentz generators ${\cal S}^{ab}$  by the 
$\tilde{S}^{ab}$ say ( or the $\tilde{\tilde{S}}{ }^{ab}$ if we 
choose them instead), see equations (\ref{eq29}, \ref{eq31}). 
This indeed means that one of the two kinds of operators 
fulfilling the Clifford  
algebra and anticommuting with the other kind - it has been made a 
choice of  $dx^a 
\tilde{\tilde{\vee}} $ in the K\" ahler case and  
$\tilde{\tilde{a}}{ }^a $ in the approach of one of us - are put 
to 
zero in the operators of Lorentz transformations; as well as in 
all the operators representing the physical quantities. The  use 
of $\;dx^0 \tilde{\tilde{\vee}}\;$ or $\; \tilde{\tilde{a}}{ 
}^0\; $ in  
the operator $\tilde{\gamma}^0$ is the exception  only used to 
simulate the Grassmann even  parity operation $\vec{dx}^a \to -\vec{dx}^a $ 
and $ \vec{\theta} \to - \vec{\theta}, $ respectively. 
 
The assumption, which we call ''cheat'' was made in the  
K\" ahler approach~\cite{kah} and in its lattice version~ 
\cite{bj}, as well as in the  
approach of one of us~\cite{norma}. 
 
In ref.~\cite{norma} the $\tilde{\tilde{a}}{ }^a$'s are argued away 
on the ground that with a certain single particle action  
\begin{equation} 
I = \int d \tau d \xi \quad L(x, \theta, \tau, \xi), 
\end{equation} 
(with $x^a$ being ordinary coordinates, $\theta^a$ Grassmann 
coordinates, $a \in \{0,1,..,d\}$, $\tau$ an ordinary time 
parameter and $\xi$ an anticommuting time parameter and assuming 
$X^a = x^a + \epsilon \xi \theta^a$ and making a choice for $\epsilon$)  
with which we shall not go in details here, the  
$\tilde{\tilde{a}}{ }^a$ appear to be  
zero as one of the 
constraints.  
This constraint has been used to put $\tilde{\tilde{a}}^a$'s 
equal zero in the further calculations in this reference and it 
was used as argument for dropping the 
$\tilde{\tilde{S}}^{ab}$-part of the Lorentz generator ${\cal 
S}^{ab}$.  
\footnote{ We point out  in ref.(\cite{norma}) that this 
constraint, when once being taken   
into account by putting it zero in all the physical operators, 
was not further treated as a weak equation. Furthermore such a weak 
equation - $\tilde{\tilde{a}}{ }^a$ is an odd Grassmann operator 
- can not at all be fulfilled.} 
Let us  
stress that once the $\tilde{\tilde{a}}{ }^a$ or $ dx^a 
\tilde{\tilde{\vee}}$ is  
dropped and accordingly {\em the $\tilde{\tilde{S}}{ }^{ab}$ is 
dropped} - for whatever reason - {\em one is no longer asking  
for the representation under the same Lorentz transformations  
( including rotations ) and one shall not expect to find  
say integer spin even if the field considered is purely 
constructed from scalars, vectors and tensors }! 
 
Let us    
point out further that what happens is that as well 
the $\theta^a$ polynomials of one of us as the linear 
combinations of p-forms in the K\" ahler approach can be 
formulated as {\em double spinors}, i.e. expressions with 
two (Dirac) spinor indices, $\alpha$ and $\beta$ say, 
and that the ''cheat'' consists in {\em  dropping} from the  
concept of Lorentz transformations the transformations in 
{\em one of these indices}. In fact we can rewrite:

For the even d case one has 
\begin{eqnarray} 
&{\rm either}& \quad \psi_{\alpha\beta}(\{ \theta^a \}) := \sum_{i=0}^d  
\; (\gamma_{a_1}\gamma_{a_2} \cdots \gamma_{a_i})_{\alpha\beta} 
\; \theta^{a_1}\theta^{a_2} \cdots \theta^{a_i}, 
\nonumber 
\end{eqnarray} 
\begin{eqnarray} 
& {\rm or}& \quad 
\psi_{\alpha\beta}(\{ dx^a \}) := \sum_{i=0}^d  
\; (\gamma_{a_1}\gamma_{a_2} \cdots \gamma_{a_i})_{\alpha\beta} 
dx^{a_1} \wedge dx^{a_2} \wedge \cdots dx^{a_i} \wedge, 
\label{gamae} 
\end{eqnarray} 
 
\noindent 
while for the odd d case one has: 
\begin{eqnarray} 
&{\rm either}& \quad \psi_{\alpha\beta\Gamma}(\{ \theta^a \}) := 
\sum_{i=0}^d  
\; (\gamma_{(\Gamma)a_1}\gamma_{(\Gamma)a_2} \cdots 
\gamma_{(\Gamma)a_i})_{\alpha\beta}  
\; \theta^{a_1}\theta^{a_2} \cdots \theta^{a_i}, 
\nonumber 
\end{eqnarray} 
\begin{eqnarray} 
&{\rm or}& \quad 
\psi_{\alpha\beta\Gamma}(\{ \theta^a \}) := \sum_{i=0}^d 
\; (\gamma_{(\Gamma)a_1}\gamma_{(\Gamma)a_2} \cdots 
\gamma_{(\Gamma)a_i})_{\alpha\beta}  
\; dx^{a_1}\; \wedge dx^{a_2}\; \wedge  \cdots dx^{a_i}\; \wedge, 
\label{gamao} 
\end{eqnarray} 
with the convention 
$ a_1 < a_2 < a_3<...< a_i.$  
Here the sums run over the number $i$ of factors in the products 
of $dx^a \wedge $ or $ \theta^a$ coordinates, a number, which is 
the same as the  
number of gamma-matrix factors and it should be remarked that we 
include the possibility $i=0$ which means no factors and is 
taken to mean that the product of zero $ dx^a \wedge $ or 
$\theta^a$-factors is  
unity and the product of zero gamma matrices is the unit matrix.  
 The indices $\alpha,\;\beta$ are the spinor 
indices and taking the product of gamma-matrices conceived of  
as matrices the symbol $(...)_{\alpha\beta}$ stands for an element 
in the $\alpha$-th row and in the $\beta$-th column. There is an  
understood Einstein convention summation over the contracted 
vector indices $a_k$, k=1,2,...,i. The gamma-matrices are in the 
even dimension case $2^{d/2}$ by $2^{d/2}$ matrices and in the 
odd dimension case $2^{(d-1)/2}$ by $2^{(d-1)/2}$ matrices. In 
the odd case we have worked with two (slightly) different  
gamma-matrix choices - and thus have written the gamma-matrices  
as depending on the sign $\Gamma$ as $\gamma(\Gamma)_{a_k}$ - 
namely gamma matrix choices obeying  
 
\begin{equation} 
\Gamma = \gamma_1 \gamma_2 \cdots \gamma_d. 
\end{equation} 
 
The $\gamma_a$ matrices should be constructed  
of course so that they obey the Clifford algebra  
\begin{equation} 
\{\gamma_a,\gamma_b\} = 2\eta^{ab} 
\label{antic} 
\end{equation} 
and we could e.g. choose 
 
$$\begin{array}{ccccccccccc} 
\gamma_1& :=& i\sigma_2^1 &\times & \sigma_3^2 &\times & 
\sigma_3^3& \times &\cdots&  
\times &\sigma_3^n\\  
\gamma_2 &:=& -i\sigma_1^1&\times &\sigma_3^2&\times & \sigma_3^3 
&\times &\cdots&  
\times& \sigma_3^n\\  
\gamma_3 &:=& iI^1&\times & \sigma_2^2&\times & \sigma_3^3&\times & 
\cdots&  
\times &\sigma_3^n\\  
\gamma_4 &:=& iI^1&\times & (-)\sigma_1^2&\times & \sigma_3^3&\times 
& \cdots&  
\times& \sigma_3^n\\  
\gamma_5 &:=& iI^1&\times & I^2&\times &\sigma_2^3&\times 
 & \cdots &  
\times& \sigma_3^n\\  
\vdots & & \vdots & \vdots & \vdots& \ddots & \vdots \\   
\gamma_{2n-1} &:=&iI^1&\times & I^2&\times & I^3& \times & \cdots & 
\times &\sigma_2^n\\  
\gamma_{2n} &:=&iI^1&\times  &I^2&\times & I^3&\times & \cdots&  
\times &(-)\sigma_1^n\\  
\end{array}$$ 
for an even dimension $d=2n$, while for an odd dimension 
$d=2n+1$ the gamma matrix $\gamma^{2n+1}$ has to be included  
$$\begin{array}{ccccccccccc} 
\gamma_{2n+1}& := &i\Gamma 
\sigma_3^1&\times & \sigma_3^2&\times &\sigma_3^3&\times &\cdots&\times& 
\sigma_3^n,\\  
\end{array}$$ 
with 
$\Gamma = \prod_{a}^{2n+1} \gamma_a$. 
( see e.g. (\cite{geor})). 
The above metric is supposed to be Euclidean. For the Minkowski 
metric $ \gamma_1 \rightarrow -i\gamma_1$ has to be taken, if the 
index $1$ is recognized as the "time" index. 
We shall make use of the Minkowski metric, counting the 
$\gamma^a$ from $0,1,2,3,5,..d$, and assuming the metric 
$\eta^{ab} = diag(1,-1,-1,...,-1)$. 
 
In this notation we can see that for fixed values of the  
index $\beta$ we obtain one of the four bispinors in Table I. 
conceived  of as a spinor in the index $\alpha$ and with the 
understanding that the $\tilde{a}^a$ in the table lead to  
the corresponding $\theta^a$, when acting on the vacuum state. 
The equivalent table for the K\" ahler approach follows by replacing $ 
\theta^a $ by  
$dx^a \wedge $.

It is our main point to show that the action by the operators  
$dx^a \vee $ or $\tilde{a}^a$ and $\tilde{\tilde{\vee}}$ or  
$\tilde{\tilde{a}}{ }^a$ in the representation  
based on the basis $\psi_{\alpha \beta}(\{ dx^a\})$ or 
$\psi_{\alpha \beta}(\{ \theta^a \})$  
with $\alpha, \beta  
\in \{ 1, 2, ..., \},\\ 
 \left\{ \begin{array}{rr} 
2^{(d-1)/2} \quad {\rm for} \quad d\quad {\rm odd} \\ 2^{d/2} 
\quad {\rm for} \quad d \quad 
{\rm even} \end{array}\right\}$ transforms the index $\alpha$ 
and $\beta$, respectively, of the basis $\psi_{\alpha\beta}(\{ 
dx^a \})$ or equivalently    $\psi_{\alpha\beta}(\{ 
\theta^a \})$ as follows: 
 
\begin{eqnarray} 
&{\rm either }&\quad dx^a \; \tilde{\vee}\; 
\psi_{\alpha\beta(\Gamma)}(\{  
dx^a \})\; \propto\;  
\gamma^a_{\alpha \gamma}\; \psi_{\gamma\beta(\Gamma)}(\{ dx^a 
\}), 
\nonumber 
\end{eqnarray} 
\begin{eqnarray} 
&{\rm corresponding \; to}& \quad 
\tilde{a}^a \; \psi_{\alpha\beta(\Gamma)}(\{ \theta^a \})\; 
\propto\;  
\gamma^a_{\alpha \gamma} \; \psi_{\gamma\beta(\Gamma)}(\{ \theta^a 
\}),  
\nonumber 
\end{eqnarray}

\begin{eqnarray} 
&{\rm or}& \quad 
dx^a \; \tilde{\tilde{\vee}} \; \psi_{\alpha\beta(\Gamma)}(\{ 
dx^a \})\; \propto \; 
\psi_{\alpha\gamma(-\Gamma)}(\{ dx^a \}) \; \gamma^a_{\gamma 
\beta}, 
\nonumber 
\end{eqnarray} 
\begin{eqnarray} 
 &{\rm corresponding\; to}& \quad 
\tilde{\tilde{a}}^a \; \psi_{\alpha\beta(\Gamma)}(\{ \theta^a 
\})\; \propto \; 
\psi_{\alpha\gamma(-\Gamma)}(\{ \theta^a \}) \; \gamma^a_{\gamma 
\beta},  
\label{tta} 
\end{eqnarray} 
 
\noindent 
which demonstrates the similarities between the spinors of the 
one of us approach and the K\" ahler approach: The operators 
$dx^a \; \tilde{\vee}$ and  
$\tilde{a}^a$ transform the left index of the 
basis $\psi_{\alpha\beta(\Gamma)}(\{dx^a \}) $, or 
correspondingly of the basis $\psi_{\alpha\beta(\Gamma)}(\{\theta^a 
\}) $,  
while keeping the right  
index fixed and the operators $ dx^a \tilde{\tilde{\vee}}$ and  
$\tilde{\tilde{a}}^a$ transform the right index of the basis 
$\psi_{\alpha\beta(\Gamma)}(\{dx^a  
\}) $, or correspondingly of the basis  
$\psi_{\alpha\beta(\Gamma)}(\{ \theta^a \}) $ and keep the left 
index fixed. Under the action of either $ dx^a \tilde{\vee}$ and 
$\tilde{a}^a$ or $dx^a \tilde{\tilde{\vee}}$ and  
$\tilde{\tilde{a}}^a$ the basic functions  
transform  as spinors. The index in parentheses $(\Gamma)$ is defined  
for only odd d. 
We can count that the number of spinors is $2^d$ either in the 
Manko\v c's approach or in the K\" ahler's approach; the d 
dimensional Grassmann space or the space of p-forms has $2^d$  
basic functions.

We shall prove the above formulas for action of the 
$\tilde{a}^a$ and $\tilde{\tilde{a}}^a$. The proof is also valid 
for the K\" ahler case if $\tilde{a}^a$ is replaced by $dx^a \; 
\tilde{\vee}$ and $\tilde{\tilde{a}}^a$ by $ dx^a 
\tilde{\tilde{\vee}}$.

\subsection{Proof of our formula for  action of  
$\tilde{a}^a$ and $\tilde{\tilde{a}}^a$ } 
Let us first introduce the notation 
\begin{equation} 
\gamma^A :=\; \gamma^a\gamma^b \cdots \gamma^c,\qquad 
\gamma^{\overline{A}} := \; \gamma^c\gamma^b \cdots \gamma^a, 
\end{equation} 
with $a < b < \cdots < c \in A.$ 
We  recognize that  
\begin{equation} 
{\rm Trace} \; (\gamma_A \gamma^{\overline{B}}) = {\rm Trace (I)} 
\; \delta_A{ }^B,\;\;\; 
\sum_A (\gamma_A)_{\alpha \beta} (\gamma^{\overline{A}})_{\gamma 
\delta} = {\rm Trace(I)}\; \delta_{\alpha \gamma} 
\; \delta_{\beta \delta}  
\end{equation} 
and 
\begin{equation} 
\sum_i (\gamma_{A_i})_{\alpha \beta} (\gamma^c 
\gamma^{\overline{A}_i})_{\gamma \delta} =  
{\rm Trace (I)}\;(\gamma^c )_{\alpha 
\delta} \delta_{\beta \gamma},\;\;\;\;  
\sum_i (\gamma_{A_i})_{\alpha \beta} ( 
\gamma^{\overline{A}_i}(-1)^i \gamma^c)_{\gamma \delta} = 
{\rm Trace (I)}\; (\gamma^c )_{\gamma  
\beta} \delta_{\delta \alpha}. 
\end{equation} 
Using the first equation we find 
\begin{equation} 
\theta^{A} = \frac{1}{{\rm Trace(I)}}\; 
(\gamma^{\overline{A}})_{\alpha \beta}\; \psi_{\beta \alpha 
(\Gamma)}(\{\theta^a \}). 
\end{equation} 
The index $(\Gamma)$ has the meaning for only an odd $d$. That is 
why we put it in parenthesis. 
We may accordingly write 
\begin{equation} 
\psi_{\alpha\beta(\Gamma)}(\{ \theta^a \}) :=   \sum_{i=0} 
\frac{1}{{\rm Trace(I)}}\; (\gamma_{A_i})_{\alpha \beta} 
(\gamma^{\overline{A}_i} )_{\gamma \delta} \psi_{\delta 
\gamma (\Gamma)}(\{\theta^a \}),\\ 
\end{equation} 
 with $  a_1< a_2,\cdots,< a_i \in A_i$  in ascending order and with 
$\overline{A}_i $ in descending order.

Then we find, taking into account that $\tilde{a}^a |\; > = 
\theta^a$,  $\tilde{\tilde{a}}{ }^a |\; > = -i \theta^a$, where 
$|\; >$ is a vacuum state and Eq.(\ref{eq7}) 
\begin{eqnarray} 
\tilde{a}^c \; \psi_{\alpha\beta(\Gamma)}(\{ \theta^a \}) := \; 
\sum_{i}(\gamma_{A_i})_{\alpha \beta} \tilde{a}^c \theta^{A_i} = 
\sum_{i}(\gamma_{A_i})_{\alpha \beta} \tilde{a}^c 
\tilde{a}^{A_i} |\; > =  
\nonumber 
\end{eqnarray} 
\begin{eqnarray} 
\sum_{i} 
\frac{1}{{\rm Trace(I)}}\; (\gamma_{A_i})_{\alpha \beta} 
(\gamma^c \gamma^{\overline{A}_i} )_{\gamma \delta} \psi_{\delta 
\gamma (\Gamma)} (\{\theta^a \}). 
\nonumber 
\end{eqnarray} 
 
Using  the above relations we further find 
\begin{equation} 
\tilde{a}^c \;  \psi_{\alpha \beta (\Gamma)}(\{ \theta^a \}) := 
(-1)^{\tilde{f} (d,c) } (\gamma^c)_{\alpha \gamma} 
 \psi_{\gamma  
\beta (\Gamma)} (\{ \theta^a \}), 
\label{t} 
\end{equation} 
where $ (-1)^{\tilde{f} (d,c) }$ is $\; \pm 1$, which depends on 
the operator $\tilde{a}^c$ and the dimension of the space. 
 
We find in a similar way 
\begin{eqnarray} 
\tilde{\tilde{a}}^c \; \psi_{\alpha\beta(\Gamma)}(\{ \theta^a 
\}) := \;  
\sum_{i}(\gamma_{A_i})_{\alpha \beta} \tilde{\tilde{a}}^c \tilde{a}^{A_i} 
|\; >\; = \; 
\sum_{i} (-1)^i (\gamma_{A_i})_{\alpha \beta} \tilde{a}^{A_i} 
\tilde{\tilde{a}}^c | \;>\; =  
\nonumber 
\end{eqnarray} 
\begin{eqnarray} 
\sum_{i} (-1)^i (\gamma_{A_i})_{\alpha \beta} \tilde{a}^{A_i} 
\tilde{\tilde{a}}{ }^c |\; > \; = \; 
\sum_{i} \frac{(-1)^i} 
{{\rm Trace(I)}} \; (\gamma_{A_i})_{\alpha \beta} 
( \gamma^{\overline{A}_i} \gamma^c  )_{\gamma \delta}\; \psi_{\delta 
\gamma (\Gamma)} (\{\theta^a \}), 
\nonumber 
\end{eqnarray} 
which  finally gives 
\begin{equation} 
\tilde{\tilde{a}}^c \; \psi_{\alpha\beta(\Gamma)}(\{ \theta^a 
\}) := \;  
(-1)^{\tilde{\tilde{f}}(d,c)} 
\psi_{\alpha 
\gamma(-\Gamma)} (\{\theta^a \}) (\gamma^c)_{\gamma \beta},  
\label{tt} 
\end{equation} 
with the signum $ (-1)^{\tilde{\tilde{f}}(d,c)}$ depending on 
the dimension of the space and the operator 
$\tilde{\tilde{a}}^c$.  
 
We have therefore proven the two equations which determine the 
action of the operators $\tilde{a}^a$ and $\tilde{\tilde{a}}^a$ 
on the basic function $\psi_{\alpha 
\gamma(-\Gamma)} (\{\theta^a \}) $.

\section{Getting an Even Gamma Matrix } 
According to the Eqs.(\ref{t},$\;$\ref{tt})  
it is obvious that the $\gamma^a$ {\bf matrices, entering into 
the Dirac-K\" ahler approach  or one of us approach for spinors, 
have an odd  
Grassmann character} since both, $dx^a\; \tilde{\vee}$ and 
$\tilde{a}^a$ as well as $dx^a\;\tilde{\tilde{\vee}} $ and 
$\tilde{\tilde{a}}^a$,  
have  an odd Grassmann character. They therefore transform 
a Grassmann odd basic function into a Grassmann even basic 
function changing fermion fields into boson fields. It is clear 
that  
such $\gamma^a$ matrices are not appropriate to enter into the 
equations of motion and Lagrangeans for spinors.  
 
There are several possibilities to avoid this 
trouble~\cite{norma}. One of them was presented in section 4.  
If working with $dx^a\; \tilde{\vee}$ or $\tilde{a}^a$ alone, 
putting  
$dx^a\; \tilde{\tilde{\vee}}$ or $\tilde{\tilde{a}}{ }^a$ in the 
Hamiltonian, Lagrangean and all the  
operators equal to zero,   the 
$\tilde{\gamma}^a$ matrices of an even Grassmann character can 
be defined as proposed in Eq.(\ref{eq25})  
$\tilde{\gamma}^a:= i dx^0 \; \tilde{\tilde{\vee}} \; dx^a 
\tilde{\vee} $ or  
$\tilde{\gamma}^a:= i \tilde{\tilde{a}}^0 \tilde{a}^a,$ 
which fulfill the Clifford algebra 
$ 
\{\tilde{\gamma}^a, 
\tilde{\gamma}^b \} =  
2\eta^{ab}$, while as we already have said $  \tilde{{\cal S}}^{ab} =  
- \frac{i}{4} [\tilde{\gamma}^a, 
\tilde{\gamma}^b]. 
$  
We then have  
\begin{equation} 
\tilde{\gamma}^a \psi_{\alpha\beta(\Gamma)}(\{ \}) =  
\gamma^a_{\alpha \gamma} \psi_{\gamma\delta(-\Gamma)}(\{ 
\theta^a \}) \gamma^0_{\delta \beta}.  
\end{equation}  
One can  check that $\tilde{\gamma}^a$ have all the 
properties of the Dirac $\gamma^a$ matrices. 
 
(Exchanging $dx^a\; \tilde{\vee} $ or $\tilde{a}^a$ by $dx^a \; 
\tilde{\tilde{\vee}}$ or $ \tilde{\tilde{a}}{ }^a$, 
respectively, the gamma-matrices defined as 
$\tilde{\tilde{\gamma}}^a:= {\rm either}\; i\; dx^0\; 
\tilde{\vee}\; dx^a\;  
\tilde{\tilde{\vee}}\; {\rm or} \;i \; \tilde{a}^0 \tilde{\tilde{a}}^a  
$ have again all the properties of the Dirac 
$\gamma^a$ matrices.)

\section{ Discrete Symmetries} 
We shall comment in this section the discrete symmetries of 
spinors and vectors in the Hilbert space spanned over either the 
Grassmann coordinate space or  
the space of differential forms from the point of view of the one 
particle states of massless Dirac (that is the Weyl) particles. 
 
In oder to define the discrete symmetries of the Lorentz group we 
introduce the space inversion $P$ and the time inversion $T$ 
operator   
in ordinary space-time in the usual way. We shall assume the case $d=4$. 
 
\begin{equation} 
P x^a P^{-1} = x_a, \quad T x^a T^{-1} = -x_a  
\end{equation} 
with the metric $\eta^{ab}, x^a = \eta^{ab} x_b $ already 
defined in section 2.  
Since one wants the time reversal operator to leave $p^0$, that is 
the zero component of the ordinary space-time momentum operator 
($p^a$),   unchanged ($p^0 \rightarrow p^0$), while the 
space component $\vec{p}$ should change sign ($\vec{p} \rightarrow 
-\vec{p}$), one also requires 
\begin{equation} 
T i T^{-1} = -i, \quad {\rm leading \quad to} \quad T p^a T^{-1} 
= p_a.   
\end{equation}  
 
We first shall treat spinors. 
Having the representation of spinors expressed in terms of 
polynomials of  
$\theta^a$'s in Table I.,  which also  represents the corresponding 
superpositions of p-forms if $\theta^a$ is accordingly 
substituted by $dx^a \wedge$, we expect each of the four copies  
of Dirac massless spinors to transform under discrete symmetries 
of the Lorentz transformations in an usual way. 
 
The parity operator $P$ should transform left handed spinors with 
$<\Gamma^{(4)}> = -1$ to right handed spinors with 
$<\Gamma^{(4)}> = 1$,  
without changing the spin of the spinors. This is what 
$\tilde{\gamma}^0$ (Eq.(\ref{eq25})) does for any of four copies of 
the Dirac massless spinors, which are the Weyl bispinors of 
Table I., separately. 
 
The time 
 reversal operator $T$ should transform left handed 
spinors with $<\Gamma^{(4)}> = -1$ and spin $\frac{1}{2}$ to left 
handed spinors with $<\Gamma^{(4)}> = -1$   and spin $-\frac{1}{2}$, 
what the  operator  
\begin{equation} 
T = i \tau_{int} \cdot \tau_x K, \quad \tau_{int} = 
\tilde{\gamma}^1 \tilde{\gamma^3}, \quad \tau_x = diag (-1, 1, 
1, 1), \quad {\rm and} \quad K i K^{-1} = -i  
\end{equation} 
does when applied to any of four copies of the Dirac spinors of 
Table I. This transformation involves only members of the same 
copy of the Dirac bispinor.  The operators $\tilde{\gamma}^a$ 
which are defined in Eq.(\ref{eq25}), have due to the appropriate 
choice of phases of the spinors of Table I, the usual chiral 
matrix representation ( for both approaches - the K\" ahler and 
the one of us). 
 
One would, however, expect that the time and the space reversal 
operators should work in both spaces - that is in the ordinary 
space-time and in the space of either Grassmann polynomials or 
in the space of p-forms - in an equivalent way 
\begin{eqnarray} 
P x^a P^{-1} = x_a,&{ }&\quad  
T x^a T^{-1} = -x_a,  
\nonumber 
\end{eqnarray} 
\begin{eqnarray} 
 P \theta^a P^{-1} = \theta_a, \quad &{\rm 
or\; correspondingly}& \quad P  
dx^a \wedge P^{-1} = dx_a \wedge,  
\nonumber 
\end{eqnarray} 
\begin{eqnarray}  
{\rm and} \quad T \theta^a T^{-1} = -\theta_a, \quad &{\rm or\; 
correspondingly}&  
\quad T dx^a \wedge T^{-1} = -dx_a \wedge, 
\nonumber 
\end{eqnarray} 
\begin{eqnarray}  
T i T^{-1} = -i, \quad &{\rm leading\;\; to}& \quad T p^a T^{-1} = 
p_a 
\label{ctn} 
\end{eqnarray} 
and changing  equivalently the momenta  conjugate to   
coordinates in either the one of us or the K\" ahler approach. 
 
Applying the transformation $P$ of Eq.(\ref{ctn}) on any of four 
copies of the Dirac bispinors of Table I., one obtains the same 
result as in the above, that is the standard definition of 
the space-reversal operation. 
Applying the transformation $T$ of Eq.(\ref{ctn}) on, let us 
say, the first spinor of the first copy of the Dirac bispinors 
of Table I. ( that is on ${ }^1 \psi(\theta)_1$),   one obtains  
the last spinor of the last copy ( that is ${ }^8 
\psi(\theta)_2$). The left handed spinor with spin 
$-\frac{1}{2}$ transformed to the left handed spinor of spin  
$\frac{1}{2}$, just as it did under the usual time-reversal 
transformation, except that in this case the copy of spinors has 
been changed.

One can write down the matrix representation for this second kind of 
the time-reversal transformation. If we choose for the basis the 
first copy of bispinors of Table I. and the  
fourth copy of bispinors of Table I, we obtain the matrix: 
\begin{equation} 
 T' = \left( \begin{array}{cccc} 
0 & 0 & i \sigma_2 \exp{i\varphi} K_{\varphi} & 0 \\ 
0 & 0 & 0 & i\sigma_2 \exp{i\varphi} K_{\varphi} \\ 
-i \sigma_2 \exp{-i\varphi} K_{\varphi} & 0 & 0 & 0\\ 
0 & -i \sigma_2 \exp{-i\varphi} K_{\varphi} & 0 & 0 
\end{array} \right), 
\end{equation} 
where the $\exp{i\varphi} = 1$,  due to the choice of the phase 
of the spinors in Table I., and $K_{\varphi}$ means that the 
complex conjugation has to be performed on the phase 
coefficients only, which in our case have again been chosen to be one. 
 
This is the time-reversal operation discussed  
by Weinberg~\cite{wein} in Appendix C of the Weinberg's 
book\footnote[3]{The two kinds of the time reversal operators 
has already been discussed in refs.(~\cite{norma}). The appearance 
of the second kind of the time reversal operator in the Weinberg's 
book as well as in the Wigner's book~\cite{wigner} was pointed 
out~\cite{recai} to the authors in the Workshop ''What comes 
beyond the Standard model'' at  
Bled 1999, when it was suggested that the second kind could generate 
states, which may be used to 
describe the sterile neutrinos.}. 
 
When vectors and scalars are treated in the similar way for 
either of the two approaches, it turns out that the 
time-reversal operators do not transform one copy into another 
one.

We payed attention in this section on only spin degrees of freedom. The 
complex conjugation affects, of course, the higher part of the 
internal space as well, affecting the charges of spinors, 
vectors and tensors, if one thinks of the extension~\cite{norma} 
as discussed in section 9.

\section{ Unavoidability of  Families} 
We want to look at the funny shift of the spin compared to 
the a priori spin for a field by shifting a priori generators  
$M^{ab} = L^{ab} + {\cal S}^{ab}$ out by anther set $ \tilde{M}^{ab}= 
L^{ab} +\tilde{S}^{ab}$ as a general nice idea. 
A prerequisite for that working is that the difference between 
the two proposals for Lorentz generators  
\begin{equation} 
\tilde{\tilde{M}}{ }^{ab}:= M^{ab} - \tilde{M}^{ab} 
\end{equation} 
is also a conserved set of quantities. 
In the notation above of course we find  
\begin{equation} 
\tilde{\tilde{M}}{ }^{ab} = \tilde{\tilde{S}}{ }^{ab}. 
\end{equation} 
 
Assuming that there is indeed such two Lorentz generator 
symmetries in a model, we can ask for the representation under 
both for a given set of fields, and we can even ask for 
representation under the difference algebra 
$\tilde{\tilde{M}}{ }^{ab}$. In order to shift in going from  
$M^{ab}$ to  
$\tilde{M}^{ab}$ from integer spin to half integer spin 
the representation for the fields in question must at least  
be spin 1/2 for $\tilde{\tilde{M}}{ }^{ab}$. Actually in the cases 
we discussed the $\tilde{\tilde{M}}{ }^{ab}$ were in the Dirac  
spinor representation. But that means that the representation 
of the fields which shift representation going from  
$M^{ab}$ to $\tilde{M}^{ab}$ have to belong under 
$\tilde{\tilde{M}}{ }^{ab}$ to at least a spin 1/2 which means at 
least the Weyl spin representation of the Lorentz group, and 
that has $2^{(d/2 -1)}$ dimensions.  
But that means then that a given representation of the  
final $\tilde{M}^{ab}$ Lorentz group always must occur in 
at least $2^{(d/2-1)}$ families. 
 

\section{Generalization to Extra Dimensions}  
We have discussed the connection between the Grassmann 
$\theta^a$ formulation and the K\" ahler formalism for general 
dimension $d$ and thus we could apply it simply in the 
d=4 case, or we could use it in extended models with extra 
dimensions. One should note that the connection between the 
spinor and the forms is such that for each extra two dimensions 
the number of components of a Dirac-spinor goes up by a factor 2, 
and at the same time the number of families also doubles. 
This agrees with that adding one extra $\theta^a$ doubles the  
number of terms in the $\theta^a$ polynomials and thus adding 
two would make this number four times as big. 
 
Let us now study the application of the extra degrees of freedom 
which consists in supposing the K\" ahler degrees of freedom  
or equivalently the Grassmann $\theta^a$'s we discussed 
to the case where the $d$ dimensional space is used in a 
Kaluza-Klein type model. That is to say we here look at a 
Kaluza-Klein model extended with $\theta^a$'s or the forms, 
much more rich than usual Kaluza-Klein. It has long been 
suggested~\cite{norma} that special kinds of rotations  
of the spins especially in the extra $(d-4)$ dimensions manifest 
themselves as generators for charges observable at the end for 
the four dimensional particles. Since both the 
extra dimension spin degrees of freedom and the ordinary spin 
degrees of freedom originate from the $\theta^a$'s or the forms 
we have a unification of these internal degrees of freedom. 
We can say then that the generators rotating these degrees of 
freedom, namely the just mentioned charges acting as hinger 
dimensional spins (at high energy) and the 4-dimensional spin, 
are unified. 
 
Such rotations of the internal spin degrees of freedom  
would in order to correspond to a Kaluza-Klein gauge   
fields with massless gauge bosons have to represent full  
symmetries of the vacuum state, i.e. they should as in  
usual Kaluza-Klein correspond to Killing-vectors, but 
with the further degrees of freedom also corresponding to  
symmetry for the latter. So at the end we may consider  
also the charges associated with the internal spin as 
ordinary Klauza-Klein charges, of course in the sense of being 
for the very rich model considered here. But of course  
unless we have the $\theta^a$ or forms degrees of freedom 
one could risk that the gauge field from such symmetry could be  
practically decoupled.    
 
Let us now look at what the ''families'' found in the  
Dirac-K\" ahler will develop into in case we use it for 
a Kaluza-Klein type model, as just proposed: Usually the number of  
surviving massless fermions into the (3+1) space consists  
only of those which are connected with ''zero-modes''. 
This is to be understood so that we imagine Weyl particles  
in the high ($d$) dimensional space because of an Atiah-Singer 
theorem in $(d-4)$ dimensional ''staying compactified'' space 
ensures some modes with the extra dimension part of the  
Dirac operator gets zero for some number of modes - for  
each $d$-dimensional family. 
 
 If the model had a strength for compact space Atiah-Singer 
theorem $''A.S.strenght''$ and if the dimension of the  
full space, the number of $\theta^a$'s, is $d$, so that the 
number of families at the $d$-dimensional level becomes  
$2^{d/2}$, the total number of at low energy observable  
''families '' should be 
\begin{equation} 
\# families = ''A.S.strength'' * 2^{d/2}.  
\end{equation} 
 
As an example take the model~\cite{norma} which has $d=14$ and at 
first - at the high energy 
level - $SO(1,13)$ Lorentz group, but which should be broken to 
( in two steps ) to first $SO(1,7)\times SO(6)$ and then to 
$SO(1,3)\times SU(3)\times SU(2) $.

\section{Discussion of Species Doubling Problem} 
We may see the appearance of equally many ( namely $2^{d/2-1}$)  
right handed and left handed ''flavours '' in the K\" ahler 
 model as an expression for the no go theorem\cite{nn} for putting  
chiral charge conserving fermions on the lattice in as far as  
we could make attempts to make lattice fermions along the lines  
of Becher and Joos~\cite{bj} 
. In fact it would of course have been a  
counterexample to the no go~\cite{nn} theorem if there had been a 
different number of  right   and of left Weyl particle species 
in the Becher-Joos model, because in the free model the number 
of particles  
functions as a conserved charge.  
 
As is very well known the  
Becher-Joos model really is just the Kogut-Susskind~\cite{ks} lattice 
fermion model, it is also well known that it does not violate  
the no go theorem~\cite{nn} and this is because there is this species 
doubling, which can be interpreted as the flavours. 
 
Becher and Joos show that the Kogut-Susskind lattice 
description of Dirac fields is equivalent to the lattice 
approximation of the Dirac-K\" ahler equation. 
( see page 344 in the Becher-Joos\cite{bj} article). 
 
This Kogut-Susskind model is one that gives us Dirac 
particles, but we can seek to get to Weyl particles in 
a naive $\Gamma^{(4)}$ ( or 
$\tilde{\Gamma}^{(4)} $ or $\gamma^5$ in the usual notation ) 
projecting way, but of course now such a  
projection would have to be translated into the language  
with the vector and scalar fields in the K\" ahler's formulation, 
and it is rather easy to see~\cite{Bled5c11}  
that requiring only one $\Gamma^{(4)}$  projection implies that 
the coefficient to one p-form say $dx^H$ should relate 
( just by a sign $\times i$) to that of the by the Hodge star $*$ 
associated $*dx^H$ (See subsection 4.5). Actually we easily see 
that requiring  
the restriction that  
\begin{equation} 
(1 + \Gamma^{(4)}) \psi = 0 
\end{equation} 
in the language of K\" ahler becomes  
\begin{equation} 
(1 + i*)u=0. 
\end{equation} 
 
If we want like Joos and Becher to put the theory on 
the lattice there is a difficulty in just imposing this  
constraint, because the natural relation imposed by the  
Hodge star $*$ on the lattice would go from lattice to the dual 
lattice and we could not identify without a somewhat ambiguous 
choice the $*$dual of a given lattice element, so as to 
impose the ''self duality'' condition.

Could we possibly invent a way to circumvent the no go 
theorem~\cite{nn}  
for chirality conserving fermions on the lattice by making  
the species doublers bosons instead of fermions, both having 
though spin 1/2 ? 
 
In the formulation by one of us which we have related to  
the K\" ahler formulation there is ( naturally) assigned 
different Grassmanian character to different components of the 
wave function. In fact the wave function with coefficients to  
monomial terms that are products of different sets of  
( mutually different) $\theta^a$-variables - in the sense of 
course that a polynomial is given by its coefficients -, 
and thus the coefficients to the products with an even number  
of factors have different Grassmannian character from those  
of the odd number of factors. That actually is in the theory 
of one of us somewhat of an embarrassing reason for a super selection 
rule, which though may be overcome by taking into account the  
charges related to extra dimensions appearing in that model. 
But here we now want to point out the hope that these very  
Grassmann character problems may be used as a new idea to 
circumvent the no go theorem. In fact we could hope for  
 that spin $1/2$ and say left handed 
flavour appear with fermionic statistics (the Grassmann odd 
character) while  spin $1/2$ flavour with bosonic 
statistics would appear as right handed, and that even on the 
lattice.

\section{Concluding Remarks} 
The way that Manko\v c~\cite{norma} chooses to quantize the system,  
that is a particle moving in ordinary and Grassmann 
coordinate space, 
is to let the wave function be allowed to be any function of the 
$d$ Gassmann variables $\theta^a$, so that any such function 
represents a state of the system. But in this quantization the 
$\tilde{\tilde{a}}{ }^a$'s can not be put weakly to  zero.  
In other words that quantization turned out not to obey the 
equation expected from expression for the canonical momentum  
$p^{\theta a}$, being proportional to the coordinate $\theta^a$ 
as derived from the Lagrangian.  
If, however, in the operators such 
as the Hamiltonian and the Lorentz transformation operators  
$\tilde{\tilde{a}}{ }^a$'s are just put strongly to zero, so 
that all the  
operators only depend on $\tilde{a}^a$, while either 
$\tilde{a}^a $ or $ \tilde{\tilde{a}}{ }^b$ fulfill the Clifford 
algebra: $\{\tilde{a}^a,  
\tilde{\tilde{a}}{ }^b \} = 0$ and $\{ \tilde{a}^a, \tilde{a}^b \} 
= 2 \eta^{ab} = \{ \tilde{\tilde{a}}{ }^a, \tilde{\tilde{a}}{ }^b 
\} $, 
the expressions obtained after having put the 
$\tilde{\tilde{a}}{ }^a$'s to zero describe spinor degrees of 
freedom. In particular, only  
the operators $\tilde{S}^{ab}$ are used as the Lorentz 
generator. One has accordingly  
the new Lorentz transformations  instead of the  
a priori one in the wave function on Grassmann space quantization 
used. In that case  the  
argument for only having integer spin  breaks down, what 
 the calculations indeed confirm  to happen. 
 
We should now attempt to get an understanding of what goes on 
here by using a basis inspired from the 
Dirac-K\" ahler-construction, which is a way often used on lattices 
to implement fermions on the lattice. The Dirac-K\" ahler 
construction starts from a field theory with a series of  
fields which are 0-form, 1-form, 2-form, ...,d-form. 
They can be thought of as being expanded on a basis of all the 
wedge product combinations of the basis $dx^1$, $dx^2$, 
...,$dx^d$ for the one-forms, including wedge products from zero 
factors to d factors. In the Dirac-K\" ahler construction one 
succeeds in constructing out of these ``all types of forms''  
$2^{d/2}$ Dirac spinor fields. This construction is without 
a ''cheat'' impossible in much the same way as Manko\v c's 
approach  ought to be.  
 
We have pointed out clearly in this paper how this ''cheat'' 
occurs in both approaches, showing up  all the 
similarities of the two approaches and using the simple 
presentation of the quantum mechanics in Grassmann space to not 
only simplify the  
Dirac-K\" ahler approach but also to generalize it.  
We have shown in particular that in both approaches besides the 
(two kinds of )  
generators for the Lorentz transformations for spinors also the 
generators for vectors and tensors exist. There are four 
copies of the Weyl bispinors. One kind of the spinorial 
type of the Lorentz transformations defines the Weyl spinors, 
another  
kind transforms one copy of  Weyl spinors  into another of 
the same Grassmann character. We also have shown 
the two kinds 
of the time reversal operators, as well as the fact that  in   
Grassmann space or space of differential forms of d dimensions, 
$d>4$, spins and charges unify. We 
pointed out the necessity of defining the gamma-matrices of  
 an even Grassmann character.

\section{Acknowledgment. } This work was supported by Ministry of 
Science and Technology of Slovenia as well as by funds NBI - HE 
- 99 - 35, CHRX - CT94 - 0621, INTAS 93 - 3316, INTAS - RFBR 95 
- 0567.


\begin{thebibliography}{99} 
 
\bibitem{kah} E. K\" ahler, Rend. Mat. Ser. {V \bf 21}, 452  
(1962). 
 
\bibitem{norma} N. Manko\v c Bor\v stnik, Phys. Lett. {\bf B   
292}, 25 (1992); N. Cimento {\bf A 105}, 1461 (1992); 
J. Math. Phys. {\bf 
34}, 3731 (1993); Int. Jour. Mod. Phys. {\bf A 9} 1731 (1994); J. 
Math. Phys. {\bf 36}, 1593 (1995);  
Mod. Phys. Lett. {\bf A 10}, 587 (1995); hep-th9408002; 
hep-th9406083; N. Manko\v c Bor\v stnik, S. Fajfer, N. Cimento, 
{\bf 112 B} 1637(1997) 
Proceedings of the International conference quantum systems, 
New trends and methods, Minsk, 23-29 May, 1994, p. 312, Ed. by 
A.O. Barut, I.D. Feranchuk, Ya.M. Shnir, L.M. Tomil'chik, World 
Scientific, Singapore 1995;  Proceedings of the US-Polish 
Workshop physics from Plank scale to electroweak scale, Warsaw, 
21-24 Sept. 1994, p. 86, Ed. by P. Nath, T. Taylor, S. Pokorski, 
World Scientific, Singapore 1995;  Proceedings of the 
$7^{th}$ Adriatic meetings on High energy physics, Bri\-ju\-ni, 
Croatia, 13-22 Sept.1994, p. 296, Ed. D. Klabu\v car, I. Picek, 
D. Tadi\' c, World Scientific, Singapore 1995; Proceedings of 
the Barut memorial conference on Group theory in physics, Tr. J. 
of Phys.{\bf 21} 321 (1997), Proceedings to the international 
workshop on What comes beyond the Standard model, Bled, 
Slovenia, 29 June-9 July 1998, Ed. by N. Manko\v c Bor\v stnik, 
H. B. Nielsen, C. Froggatt, DMFA Zalo\v zni\v stvo 1998, p. 20; 
 Proceedings of 
the $VII^{th}$ international conference Symmetry methods in 
physics, Dubna 10-18 July 1995, p. 385, Ed. by Sissakian, G.P. 
Pogosyan, Publ. Dept., Joint Institute for Nuclear Research, 14 
19 80 Dubna, ISBN 5 85165 453 8. 
 
\bibitem{ravn} P. Di Vecchia, F. Ravndal, Phys. Lett. {\bf A 
73}, 371 (1979). 
 
\bibitem{bj} P. Becher, H. Joos, Z. Phys. {\bf C} - Part. and 
Fields {\bf 15, } 343 (1982). 
 
\bibitem{nn} H. B. Nielsen, M. Ninomija, Phys.  
Lett. {\bf B 105 } 219 (1981), Nucl. Phys.{\bf B 185}, 20 (1981). 
 
\bibitem{Bled5c11} H. B. Nielsen, N. Manko\v c Bor\v stnik, 
Proceedings to the international workshop on What comes beyond 
the Standard model, Bled, 
Slovenia, 29 June-9 July 1998, Ed. by N. Manko\v c Bor\v stnik, 
H. B. Nielsen, C. Froggatt, DMFA Zalo\v zni\v stvo 1998, p. 68, 
IJS.TP.99/17 or  NBI-HE-99-35 or CERN-TH/99-288, hep-th/9909169.
 
\bibitem{geor} H. Georgi, {\it Lie Algebra in Particle Physics} 
( Frontier in Physics, A Lecture Note and Reprint Series )  (The 
Benjamin Cumings 1982). 
 
\bibitem{wein} S. Weinberg, {\it The Quantum Theory of Fields, 
Vol. I} (Cambridge University Press 1995, Cambridge CB2 IRP). 
 
\bibitem{wigner} E. P. Wigner, {\it Group Theoretical Concepts and 
Methods in Elementary Particle Physics}, Ed. F. G\" ursey, (Gordon 
and Breach, New York 1964). 
 
\bibitem{recai} R. Erdem, Proceedings to the International 
workshop ''What beyond the Standard model'', Bled 22-31 July 1999. 
 
\bibitem{ks} J. Kogut, L. Susskind, Phys. Rev. {\bf D 13}, 1043 (1976). 
 
\end{thebibliography}
\end{document}